\documentclass[twocolumn,aps,prr,longbibliography,floatfix,showkeys,letterpaper,superscriptaddress]{revtex4-2}
\usepackage[utf8]{inputenc}
\usepackage{graphicx}%
\usepackage{dcolumn}
\usepackage{xcolor}
\usepackage{xspace}
\usepackage[pdfborder={0 0 0}]{hyperref} 
\usepackage{amsmath}
\usepackage{amssymb}
\usepackage{bm}
\usepackage{braket}
\usepackage{amsfonts}
\usepackage{dsfont}
\usepackage{float}
\usepackage[T1]{fontenc}
\usepackage[english]{babel}
\usepackage{cfr-lm}
\usepackage{upgreek} 

\usepackage{setspace} 

\usepackage{mathtools} 
\usepackage{soul}

\usepackage{letltxmacro}
\LetLtxMacro{\ORIGselectlanguage}{\selectlanguage}
\makeatletter
\DeclareRobustCommand{\selectlanguage}[1]{%
	\@ifundefined{alias@\string#1}
	{\ORIGselectlanguage{#1}}
	{\begingroup\edef\x{\endgroup
			\noexpand\ORIGselectlanguage{\@nameuse{alias@#1}}}\x}%
}
\newcommand{\definelanguagealias}[2]{%
	\@namedef{alias@#1}{#2}%
}
\makeatother
\definelanguagealias{en}{english}
\definelanguagealias{En}{english}
\definelanguagealias{de}{german}
\definelanguagealias{EN}{english}


\usepackage{tikz}
\usetikzlibrary{calc} 
\usetikzlibrary{decorations.pathreplacing} 
\usetikzlibrary{patterns} 
\usetikzlibrary{shapes.geometric} 
\usetikzlibrary{positioning} 
\usetikzlibrary{shapes, arrows} 

\begin{document}

\title{Arbitrary quantum circuits on a fully integrated two-qubit computation register for a trapped-ion quantum processor}

\author{N.~Pulido-Mateo}
\affiliation{Institut für Quantenoptik, Leibniz Universität Hannover, Welfengarten 1, 30167 Hannover, Germany}
\affiliation{Physikalisch-Technische Bundesanstalt, Bundesallee 100, 38116 Braunschweig, Germany}
\author{H.~Mendpara}
\affiliation{Institut für Quantenoptik, Leibniz Universität Hannover, Welfengarten 1, 30167 Hannover, Germany}
\affiliation{Physikalisch-Technische Bundesanstalt, Bundesallee 100, 38116 Braunschweig, Germany}	
\author{M.~Duwe}
\affiliation{Institut für Quantenoptik, Leibniz Universität Hannover, Welfengarten 1, 30167 Hannover, Germany}
\affiliation{Physikalisch-Technische Bundesanstalt, Bundesallee 100, 38116 Braunschweig, Germany}
\author{T.~Dubielzig}
\affiliation{Institut für Quantenoptik, Leibniz Universität Hannover, Welfengarten 1, 30167 Hannover, Germany}
\author{G.~Zarantonello}
\affiliation{Institut für Quantenoptik, Leibniz Universität Hannover, Welfengarten 1, 30167 Hannover, Germany}
\author{L.~Krinner}
\affiliation{Institut für Quantenoptik, Leibniz Universität Hannover, Welfengarten 1, 30167 Hannover, Germany}
\affiliation{Physikalisch-Technische Bundesanstalt, Bundesallee 100, 38116 Braunschweig, Germany}
\author{C.~Ospelkaus}
\affiliation{Institut für Quantenoptik, Leibniz Universität Hannover, Welfengarten 1, 30167 Hannover, Germany}
\affiliation{Physikalisch-Technische Bundesanstalt, Bundesallee 100, 38116 Braunschweig, Germany}
\affiliation{Laboratorium für Nano- und Quantenengineering, Leibniz Universität Hannover, Schneiderberg 39, 30167 Hannover, Germany}

\date{\today}%

\begin{abstract}
	We report on the implementation of arbitrary circuits on a universal two-qubit register that can act as the computational module in a trapped-ion quantum computer based on the quantum charge-coupled device architecture. A universal set of quantum gates is implemented on a two-ion Coulomb crystal of $^9\mathrm{Be}^+$ ions using only chip-integrated microwave addressing. Individual-ion addressing is implemented using microwave micromotion sideband transitions; we obtain upper limits on addressing cross-talk in the register. Arbitrary two-qubit operations are characterized using the cycle benchmarking protocol.    
\end{abstract}


\maketitle

Trapped ions are among the most promising platforms for building a universal, general purpose, digital quantum computer~\cite{cirac_quantum_1995,wineland_experimental_1998,kielpinski_architecture_2002, brown_co-designing_2016,bermudez_assessing_2017, bruzewicz_trapped-ion_2019}. Among the key features are the availability of long-lived (e.g.\ hyperfine) state-pairs, long coherence times~\cite{langer_long-lived_2005}, high-fidelity quantum gates~\cite{gaebler_high-fidelity_2016,ballance_high-fidelity_2016,baldwin_high-fidelity_2021, srinivas_high-fidelity_2021}, all-to-all connectivity either within one crystal~\cite{cirac_quantum_1995,postler_demonstration_2022} or enabled by transport~\cite{wineland_experimental_1998,kielpinski_architecture_2002,amini_toward_2010}. Photons may be used to generate distributed entanglement for networking of remote trapped-ion quantum computers~\cite{stephenson_high-rate_2020,krutyanskiy_entanglement_2023}. Microfabricated and surface-electrode~\cite{chiaverini_surface-electrode_2005,seidelin_microfabricated_2006,labaziewicz_suppression_2008} ion traps provide a scalable platform for the implementation of the quantum charge-coupled device (QCCD) architecture, where all necessary operations are implemented in different specialized zones, interconnected by transport of the ions. An elementary ``quantum core'' could consist of a two-qubit computation register, combined with a junction and suitable storage registers as well as an individual-ion readout and state preparation register. 

In view of scaling the trapped-ion platform, it is desirable to integrate core aspects of the control of trapped-ion qubits into a scalable microfabricated trap structure. Recent examples include trap-integrated detectors~\cite{todaro_state_2021,setzer_fluorescence_2021,reens_high-fidelity_2022} or optical addressing~\cite{mehta_integrated_2020, niffenegger_integrated_2020}. An earlier example is the two-qubit gate of~\cite{ospelkaus_microwave_2011}, based on chip-integrated microwave elements~\cite{ospelkaus_trapped-ion_2008,ospelkaus_microwave_2011}. Subsequent work on this technology demonstrated the implementation of individual-ion addressing 
~\cite{warring_individual-ion_2013}, high-fidelity two-qubit gates~\cite{harty_high-fidelity_2016,srinivas_high-fidelity_2021} and a single-port device for quantum gates~\cite{hahn_integrated_2019}, robust against different noise sources~\cite{zarantonello_robust_2019,duwe_numerical_2022}. High fidelity single-qubit gates from integrated microwave elements have also been demonstrated~\cite{brown_single-qubit_2011,harty_high-fidelity_2014,craik_high-fidelity_2017,leu_fast_2023}. In~\cite{srinivas_high-fidelity_2021}, the joint execution single-qubit and two-qubit gates was also implemented to generate specific two-qubit entangled states with record fidelity. See~\cite{mintert_ion-trap_2001} and~\cite{khromova_designer_2012} for a complementary microwave approach using static field gradients. 

In this letter, we demonstrate the execution of arbitrary two-qubit circuits on a universal two-qubit computation register. All quantum gates were executed using  on chip-integrated elements only. We place upper bounds on cross-talk infidelities in individual-ion addressing and demonstrate how to establish the proper phase relationship for combining individual-ion addressing with two-qubit gates in our approach. Finally, we show the execution of arbitrary two-qubit circuits by applying the cycle benchmarking protocol~\cite{erhard_characterizing_2019}.

We implement the two-qubit register in a single-layer micro-fabricated Paul trap~\cite{hahn_integrated_2019}. The qubits are encoded in the internal hyperfine ground states of two $^9\mathrm{Be}^+$ ions on a first-order magnetic-field independent transition, with $\left|F=1,
\,m_F=1\right>=\left|\uparrow\right>$ and $\left|F=2,\,m_F=1\right>=\left|\downarrow\right>$ at a quantization field of $\left |\mathbf{B}_0\right |\approx22.3~$mT~\cite{langer_long-lived_2005,wahnschaffe_single-ion_2017}. The register is designed for trapping ions at an approximate height of $70\,\mu\mathrm{m}$ from the surface. A total of 10 DC control electrodes provide axial confinement and enables fine adjustment of the position and orientation of a two-ion crystal, while one split RF electrode provides the radial confinement. A microwave signal applied to an integrated S-shaped microwave conductor~\cite{carsjens_surface-electrode_2014} will produce a two-dimensional quadrupole microwave field in the radial plane; the center of the quadrupole field coincides with the pseudopotential minimum of the Paul trap. 

The resulting trap frequencies are $\omega_{x,\,y,\,z}=2\pi\times\{1.10(1),\,6.20(1),\,6.25(1)\}~$MHz. We ground-state cool the ion crystal's radial lower-frequency in-phase and out-of-phase motional modes to $\bar{n}_{\text{IP}}=0.4(2),\,\bar{n}_{\text{OOP}}=0.1(2)$ while the other modes of motion are Doppler cooled only (estimated at $\bar{n}_{\text{rad}}\approx2$ and $\bar{n}_{\text{ax}}\approx10$ -- note that the axial direction can only be estimated, as we have no diagnostic for this motional direction).

The RF field oscillates at $\omega_{\text{RF}} \approx 2\pi \times 88.2$~MHz. If an ion is held at a position of non-vanishing RF field, this will result in micromotion at $\omega_{\text{RF}}$ with an amplitude $\vec{r}_{\text{MM}}$. If the microwave conductor is driven at $\omega_0\pm\omega_{\text{RF}}$, where $\hbar \omega_0=E_{\left|\uparrow\right>} - E_{\left|\downarrow\right>}$, then in the co-moving frame of the ion, it will experience an oscillating field component at $\omega_0$. As a result, the ion will undergo Rabi oscillations (micromotion sidebands~\cite{berkeland_minimization_1998}) with a Rabi rate~\cite{warring_individual-ion_2013,warring_techniques_2013}
\begin{equation}
	\Omega_{\text{MM}}=\frac{1}{2} \frac{B'}{\sqrt{2}}|\vec{r}_{\text{MM}}|\frac{\mu}{2\hbar}\ ,
\end{equation}
where $\mu$ is the transition matrix element, $2\pi\hbar$ is Planck's constant, and $B^{\prime}$ is the applied magnetic-field gradient. Individual-ion addressing can be implemented if one ion experiences strong micromotion, while the other ion ideally experiences vanishing micromotion~\cite{warring_individual-ion_2013}. 

Two different configurations of the crystal can be prepared to address each one of the ions. This is accomplished by finding two sets of DC potentials, each one twisting the crystal into the appropriate configuration. In each configuration, the ion to be hidden from the interaction sits as close as possible to the null of the radially confining RF field. Transport to and from each configuration is done adiabatically in approximately $100\,\mu\mathrm{s}$.

Fine adjustment of the crystal's final position is done via a superimposed set of DC potentials which are designed to displace the crystal in the radial plane. The calibration procedure is based on minimizing the Rabi frequency of the qubit encoded in the non-addressed ion, and is accomplished by changing the values of the DC potentials while performing a Rabi flopping experiment on the adressed ion. An attempt to fit a Rabi oscillation of the non-addressed ion yields a Rabi rate compatible with zero ($\Omega_{\mathrm{NA}}=0.01(10)~$kHz).

\begin{figure}[htb!]
	\centering
	\includegraphics[width=\columnwidth]{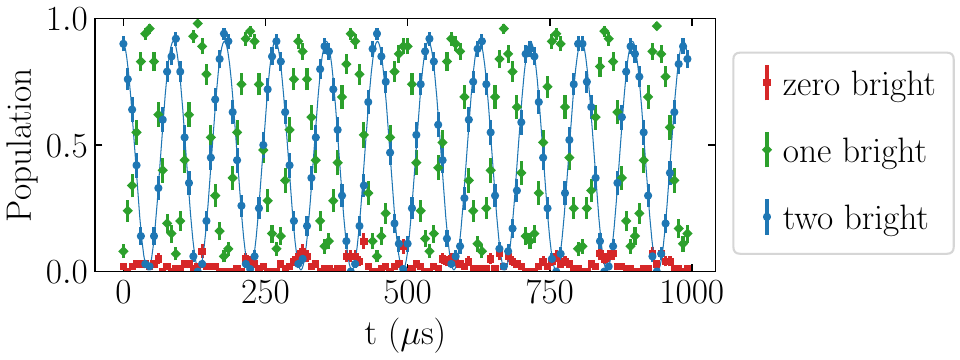}
	\caption{Rabi flopping of one ion in the first configuration of the Coulomb crystal using the micromotion sideband. The state is prepared in $\left|\uparrow\uparrow\right>$ (two-ion bright). The two-ion bright population (blue) is a proxy for the flopping of a single ion and is fitted to a sine function to extract the Rabi rate. The $\left|\downarrow\downarrow\right>$ population (red) shows slight spikes that coincide with the maxima of the one-ion bright population. This is caused by state preparation and measurement imperfections.}
	\label{fig:rabiflop}
\end{figure}

Fig.~\ref{fig:rabiflop} shows the time evolution of the internal states of a two-ion crystal, while we selectively address one of the qubits by means of a micromotion sideband. Our detection system can only register the global fluorescence of the two-ion crystal. We can therefore only distinguish between the zero-ion bright, one-ion bright and two-ion bright states. The ions are initialized such that, without any micromotion-sideband interaction, they will be detected as bright. The signature of individual-ion addressing is then given by oscillations between the two-ion bright and one-ion bright populations, while a zero-ion bright signal is ideally never detected. In the first configuration of the crystal, we measure a Rabi rate of $\Omega_{\mathrm{ion_1}}=2\pi \times 11.15(2)$~kHz. In the second configuration, we have $\Omega_{\mathrm{ion_2}}=2\pi \times 5.49(7)$~kHz. In both cases a magnetic field gradient of $B^{\prime}\approx 12~$T/m was used. We note that the large difference in the Rabi-rates is purely due to different micromotion amplitudes, as the configurations were not symmetric.

To examine potential cross-talk between the two ions, we adopt a simplified version of the technique described in Ref.~\cite{piltz_trapped-ion-based_2014}. The qubits are prepared in $\left|\uparrow\uparrow\right>$ and then random sequences of an increasing number of $\pi$-rotations (i. e. state inversions) about random axes are executed on a single (addressed) qubit. This leaves the unaddressed qubit (ideally) unchanged. After an even number of gates, the total state should have returned to the initial one. 

Based on~\cite{piltz_trapped-ion-based_2014}, the final-state fidelity of qubit $j$ with respect to $\left|\uparrow\right>$ when qubit $i$ is resonantly addressed ($j\ne i$) with $N$ pulses is given by
\begin{equation}
	\left< F_{i,j}(N)\right> = \frac{1}{2} \left( 1 +(2p_0-1) e^{-2C_{i,j}N} \right).
	\label{eq:crosstalk}
\end{equation}
Here $p_0<1$ accounts for imperfect state preparation and detection, and $C_{i,j}$ characterizes the cross-talk per gate. 

\begin{figure}[htb!]
	\centering
	\includegraphics[width=\columnwidth]{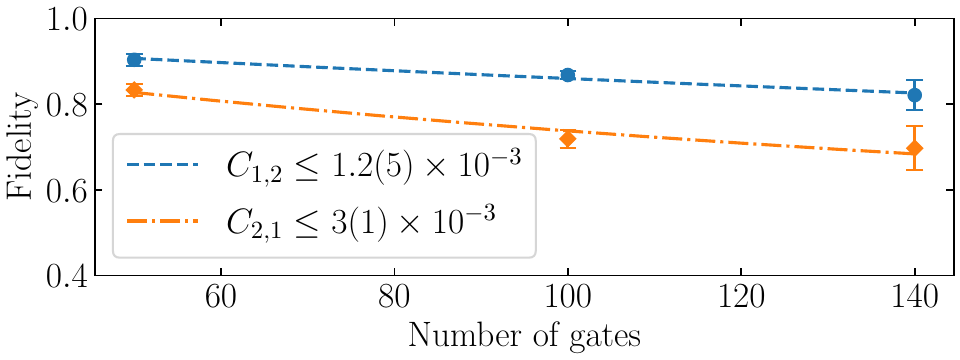}
    \caption{Cross-talk measurements for individual-ion addressed gates. The measured fidelity is given by the two-ion bright signal ($\left|\uparrow\uparrow\right>$ population). Cross-talk and $p_0$ are extracted from the fitted models. The fitted value of $p_0$ is $0.96(4)$ and $0.95(8)$ for the experimental data corresponding to the measurement of $C_{1,2}$ and $C_{2,1}$ respectively.}
	\label{fig:crosstalk}
\end{figure}

Because our detection method only allows us to distinguish between the zero-ion bright, one-ion bright and two-ion bright cases, any experimental deviation from the initial state $\left|\uparrow\uparrow\right>$ after the sequence is either an indicator of cross-talk (unintended rotations on the non-addressed ion $j$), or of imperfect single-qubit gates on the addressed ion $i$. We can obtain an upper bound on cross-talk by assigning any deviation from $\left|\uparrow\uparrow\right>$ to cross-talk only. We thus obtain upper limits of on cross-talk of $C_{1,2} \le 1.2(5) \times 10^{-3}$ and $C_{2,1} \le 3(1) \times 10^{-3}$ as shown in Fig.~\ref{fig:crosstalk}.

Additional information can be gained by repeating the individual-ion addressing experiment with just one ion trapped. Any deviation from the expected state is then due to single-qubit gate imperfections. Here, using the same analysis, we find $2.0(8)\times 10^{-3}$, which is on the same order of magnitude as the upper limits on cross-talk obtained above. Note that because of the limited set of pulses employed, neither of the above measurements are a good indicator of single-qubit gate infidelity, and only provide a lower bound. We therefore assume that imperfections in the single-qubit gates provide a good portion of the observed non-ideal measurement outcomes. In summary, the upper limit on cross-talk errors from single-ion addressed gates is on the same order of magnitude as our expected two-qubit gate error~\cite{zarantonello_robust_2019,duwe_numerical_2022} or less. 

We now discuss the set of single-qubit gates based on the above-mentioned micromotion sideband transitions. These allow us to perform rotations on the Bloch-sphere of the form:
\begin{equation}
	R_\varphi(\theta) = e^{-i\frac{\theta}{2} \big(\cos(\varphi)\sigma_x + \sin(\varphi)\sigma_y\big)},
\end{equation}
where $\theta$ is the amount of rotation and $\varphi$ is the azimuthal angle of the axis of rotation which lies on the equatorial plane of the Bloch sphere. The angle $\theta = t\frac{\pi}{t_\pi}$ is controlled via the interaction time $t$ and a correct calibration of the $\pi $-time $t_\pi = \frac{\pi}{\Omega}$, where $\Omega$ is the Rabi rate of the micromotion sideband transition. The azimuthal angle $\varphi$ is proportional to the phase $\varphi_\mathrm{DDS}$ of the direct digital synthesis (DDS) microwave drive. 

By choosing $\varphi=0$ or $\varphi=\frac{\pi}{2}$ we can implement the commonly used $R_x(\theta) = e^{-i\frac{\theta}{2}\sigma_x}$ or $R_y(\theta) = e^{-i\frac{\theta}{2}\sigma_y}$ rotations, respectively. We decompose $R_z$ gates into a sequence of $R_x$ and $R_y$ gates via a basis change: 
\begin{equation}
	R_z(\theta) = R_y(-\frac{\pi}{2}) \circ  R_x(-\theta) \circ  R_y(\frac{\pi}{2})\ . 
	\label{eq:transpile}
\end{equation}

We now move on to combining single-ion addressed gates with two-qubit gates. The entangling operation is realized by utilizing the S-shaped microwave electrode. We implement a M{\o}lmer-S{\o}rensen gate \cite{molmer_multiparticle_1999,solano_deterministic_1999,milburn_ion_2000} on the low-frequency radial out-of-phase eigenmode of the two-ion crystal, by irradiating the ion simultaneously with red and blue microwave sidebands close to the secular motional resonance. The gate sideband tones are amplitude modulated~\cite{zarantonello_robust_2019,duwe_numerical_2022} to suppress the sensitivity to motional-mode frequency fluctuations. For more details we refer the reader to the original implementations done on this apparatus ~\cite{hahn_integrated_2019, zarantonello_robust_2019}. In all the experiments in this paper we use entangling gates with a gate time of $\tau\approx 1150\,\mu\mathrm{s}$, and a detuning of $\delta\approx 1.82\,\mathrm{kHz}$. For previous implementations we were able to achieve two-qubit gate fidelities of $99.3(4)\%$\cite{duwe_numerical_2022}, as measured by partial state tomography. Earlier work for longer two-qubit gates resulted in fidelities of $99.7(1)\%$\cite{zarantonello_robust_2019}. The stroboscopic gate propagator takes the form
\begin{equation}
	U(\tau) = e^{-i\sum_{j,k=1}^2 \Phi_{jk}\sigma_{\bar\phi}^{(j)}\sigma_{\bar\phi}^{(k)}}	,
\end{equation}
where $\sigma_{\bar\phi}^{(j)} \coloneqq \frac{1}{2} \big( \sigma_x^{(j)}\cos(\bar\phi) - \sigma_y^{(j)}\sin(\bar\phi) \big)$ depends on the mean phase $\bar\phi = \frac{1}{2}\left( \phi^R + \phi^B \right)$ of the red and blue sideband drives. Gate operation on an out-of-phase mode results in a geometric phase of 
\begin{equation}
	\Phi_{jk} = 
	\begin{cases}
		+\frac{\pi}{2} \quad j=k \\
		-\frac{\pi}{2} \quad j\neq k .
	\end{cases}
\end{equation}
Without loss of generality we set the phase of the sideband DDS signal generators to zero. The two-qubit gate is then given by 
\begin{equation}
	R_{xx}(-\frac{\pi}{2}) = e^{-i\frac{\pi}{4}} \cdot e^{+i\frac{\pi}{4} \sigma_x \otimes \sigma_x}.
\end{equation}

Because the drive involved in the individual-ion addressing involves the RF drive frequency, the relative phase between the RF drive and the sideband drive matters. We cannot directly derive the RF drive frequency from our control system~\cite{langer_high_2006} because its DDS outputs are occasionally re-set, which would lead to ion loss. We therefore derive the RF signal from a phase-locked loop (PLL) circuitry implemented in the ADF4351 evaluation board from Analog Devices, which is synchronized to a DDS channel of the experimental control system. The phase relationship between the RF drive and the DDS channels used for driving the microwave electrode for quantum gates is thus reproducible from experiment to experiment. The RF drive is amplitude-stabilized~\cite{harty_high-fidelity_2013,harty_github_nodate} before being connected to a co-axial resonator. 

We measure experimentally the phase offset $\varphi_{\mathrm{offset}}$ between single-qubit gates, which includes the RF drive and the frame of reference of the two-qubit gate $R_{xx}(-\frac{\pi}{2})$, using the following operation
\begin{equation}
	\left|\psi(\varphi)\right> = \left(R_\varphi(\frac{\pi}{2}) \otimes \mathbb I\right) 
	\left(  \mathbb I \otimes R_\varphi(\frac{\pi}{2})\right)
 	R_{xx}\big(-\frac{\pi}{2}\big) 
 	\left|\uparrow\uparrow\right>
\end{equation}
while varying $\varphi$ by scanning $\varphi_{\mathrm{DDS}}$. The parity of the resulting populations will cross zero at $\varphi_{\mathrm{DDS}} = -\varphi_{\mathrm{offset}}$, with an average uncertainty of 0.85(24) degrees, and have a negative slope. 

To show the execution of arbitrary operations on two qubits, we choose the method of cycle benchmarking~\cite{erhard_characterizing_2019} to characterize our two-qubit computational module. Its results are independent of state preparation and measurement errors and the method is conceptually of simpler implementation than randomized benchmarking~\cite{gaebler_randomized_2012}. When excecuting cycle benchmarking, we obtain a composite process fidelity of $F=96.6(4)~\%$. Cycle benchmarking dresses each two-qubit operation between single-qubit operations, so this is the total fidelity of executing one two-qubit operation and one single-qubit operation on each qubit. For details, we refer the reader to the appendix and \cite{erhard_characterizing_2019}. This constitutes the key result of this publication. 

Our work thus demonstrates for the first time a universal fully chip-integrated computation register running arbitrary algorithms. By combining the computation module with transport to and from a storage register \cite{wineland_experimental_1998,kielpinski_architecture_2002}, arbitrary computations on $N$ qubits can be achieved. Ideally, laser cooling and readout would be carried out in a separate trap zone, so that residual laser-induced charging does not affect the trap potential in the computation zone and individual-ion readout can be implemented. The microwave approach employed here is quite insensitive to the initial motional state of the ion~\cite{ospelkaus_trapped-ion_2008}; yet, at some level, sympathetic cooling~\cite{kielpinski_sympathetic_2000} may become necessary. 

We now discuss the composite process fidelity obtained in this experiment. One would intuitively expect this to be dominiated by the two-qubit gate infidelity. The reason that the composite process fidelity turns out to be lower is an ac-Zeeman shift~\cite{warring_techniques_2013} introduced by the micro-motion sideband drive during the single-qubit gates. This effect was accidentally not calibrated (we properly calibrated it in the case of the two-qubit gates). This effect is discussed in detail in the appndix.

In this work, we have demonstrated the execution of arbitrary two-qubit circuits using a fully chip-integrated microwave gate mechanism in a surface-electrode ion trap. Using the cycle benchmarking protocol, we extract a composite process fidelity of 96.6(4)~$\%$, currently limited by an uncalibrated spectroscopic shift. The  register can serve as a fully integrated computation register of a QCCD-based quantum computer. The setup is currently undergoing the transition to the new experiment control system ARTIQ~\cite{bourdeauducq_artiq_2021}; once this transition is complete, we will proceed with the systematic characterization of multi-qubit quantum gate sequences with high fidelities. Individual-ion charge-coupled device (CCD) readout~\cite{halama_real-time_2022} will be available to complement individual-ion addressing; alternatively, transport can be used for individual-ion readout. In future work, we plan to extend the computation register with separate registers for storage and readout to increase the number of qubits.

We thank D.~Leibfried, D.~Slichter, P.~O.~Schmidt and F.~Wolf for helpful discussions. We are grateful to F.~Schwartau, M.~Schubert, I.~Elenskyi, M.~Schilling and the low noise electronics team at TU Braunschweig for their assistance on the electronic part of the experiment. We acknowledge help from Amado Bautista-Salvador in fabrication of the ion-trap chip and initial setup of the experiment. We are grateful for support in simulation efforts from Klemens Hammerer and Marius Schulte. We acknowledge funding from the European Union Quantum technology flagship under project `Millenion-SGA1', from `QVLS-Q1' through the VW foundation and the ministry for science and culture of Lower-Saxony, from BMBF through the project `ATIQ', from the Deutsche Forschungsgemeinschaft (DFG, German Research Foundation) under Germany’s Excellence Strategy - EXC-2123 QuantumFrontiers - 390837967 and through  the collaborative research center SFB 1227 DQ-\textit{mat}, project A01, and from PTB and LUH.

\appendix

\section{Adapted two-qubit cycle benchmarking} 

\begin{figure}[htb!]
	\centering
	\includegraphics[width=\columnwidth]{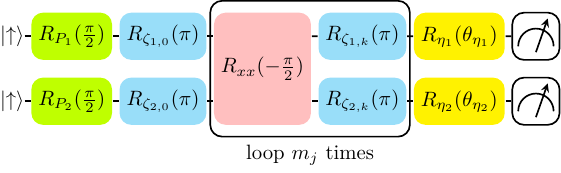}
	\caption{Schematic of the algorithm used to extract the composite process fidelity~\cite{erhard_characterizing_2019}. $R_{xx}(-\frac{\pi}{2})$ gates (pink) refer to the entangling gate. $R_{P_i}(\frac{\pi}{2})$ gates (green) are initial basis changes, $R_{\zeta_{i, k}}(\pi)$ gates (blue) are dressing cycle gates  and $R_{\eta_{i}}(\theta_{\eta_i})$ gates (yellow) are chosen such that the final state of the system is ideally $\ket{\uparrow\uparrow}$.}
	\label{fig:cbcirq}
\end{figure}

In our case the protocol described in \cite{erhard_characterizing_2019} is adapted to be used on only two qubits and running only local and two-qubit operations. We depict a schematic of the executed quantum circuits in Fig.~\ref{fig:cbcirq}, where we see two distinct types of randomizations done by local (single-qubit) gates. First, the input basis randomization of Pauli-eigenstates used for state preparation at the beginning of the measurement sequence  $\mathcal{R}_{P_i}(\pi/2)$ with $P_i\in \{x,\,y,\,z,\,I\}$. This changes the ion state away from the state-preparation and measurement basis $Z$ to a different eigenstate of the Pauli-basis. The argument $\pi/2$ indicates the pulse area needed for a resonant Rabi-rotation to achieve this basis change. We choose all possible randomizations  here. Second, there are dressing cycle randomizations $R_{\zeta_{i,\,k}}(\pi)$ (with $\zeta_{i, k} \in \{x, -x, y, -y, z, -z, I\}$ randomly chosen) which bracket the two-qubit operations. The argument is again the pulse-area for a resonant Rabi-rotations. For each possible choice of input basis randomizations, one can have a large number $L$ of dressing cycle randomizations. In \cite{erhard_characterizing_2019} $L=10$ is chosen, for simplicity we have chosen $L=1$ throughout. 

In order to differentiate between state-preparation/measurement errors and gate errors, the above sets of circuits have to be generated for two different numbers $m_1=4,\,m_2=8$ of entangling gates $R_{xx}(-\pi/2)$. The $m_i$ have to be fixed to multiples of $4$ as this ensures $R_{xx}(-\pi/2)^{m_i}=I$  

Finally, the qubits have to be restored to the measurement basis (i. e. $Z$) using rotations $R_{\eta_{i}}(\theta_{\eta_i})$. Here, the axis $\eta_i\in \{x,\,y,\,z\}$ and the angle of rotation $\theta_{\eta_i}=\pm \pi/2\}$ are calculated in order to yield $\left|\uparrow\uparrow\right>$ for ideal gates and in the absence of state preparation/measurement errors. This is possible using only local operations as an even number of $R_{xx}(-\frac{\pi}{2})$ gates returns the system into a separable state. 
 
Each of the three operations, namely addressing one ion or the other or performing the entangling gate, requires a transport operation. We minimize the number of these operations by performing as many quantum gates as possible on one qubit before addressing the other one. This is done only by interchanging commuting operations, and in order to minimize the the computational load on our experimental control software. 

After experimentally executing each one of the circuits and measuring the resulting fluorescence, the population of the chosen target final state is identified with $f_{P,m_j, l}$. Here $P \in \mathbf{P} = \big\lbrace (P_1, P_2) | P_i \in \{x, y, z, I\} \setminus {(I, I)}\big\rbrace$ is one particular combination of pairs of single-qubit initial basis changes, $m_j$ is the number of entangling gates in the circuit and $l = 1 $ indexes the dressing cycle randomization of the set of circuits.

We recover the composite process fidelity $F$ via
\begin{equation}
	F = \sum_{P\in\mathbf{P}} \frac{1}{|\mathbf{P}|}
	\left(
		\frac{
			\sum_{l=1}^L f_{p,m_2,l}
		}
		{
			\sum_{l=1}^L f_{p,m_1,l}
		}
	\right)^{\frac{1}{m_2 - m_1}}
\end{equation}

Using the simulations detailed in the following section, we estimate the effect of only utilizing a single dressing cycle randomization $l$. Compared to using more dressing cycle randomizations, for our known gate errors, we obtain an additional systematic uncertainty of $1\%$. For smaller composite process infidelities, this systematic uncertainty decreases, but always depends on the specific errors of the utilized set of quantum gates.  

\section{Single-qubit gate error simulations}

In this appendix we give a detailed analysis of the composite process fidelity based on the known spectroscopic shift during single-qubit gates. We find independently the inadvertently uncompensated ac-Zeeman shift during single-ion addressing. Subsequently, we use these independently obtained parameters to simulate the expected experimental outcome of the experimentally realized quantum circuits.    

In \cite{hahn_two-qubit_2019} (chapter 5.1.8), we have measured the spatial dependence of the magnetic field caused by the S-shaped microwave conductor. Using the known crystal configurations created by the voltages applied to the electrodes, we find all expected spectroscopic ac-Zeeman shifts $\Delta^{(k)}_{m}$, where $(k)$ references the addressed ion and $m$ indexes the ion. Since the precise relative orientation of the micromotion gradient field and the trap-rf null position is only known to within $0.6~\mu$m horizontally and  $0.04~\mu$m vertically, we find all $\Delta^{(k)}_{m}$ for a range of relative orientations. 

\begin{figure}[htb!]
	\centering
	\includegraphics[width=\columnwidth]{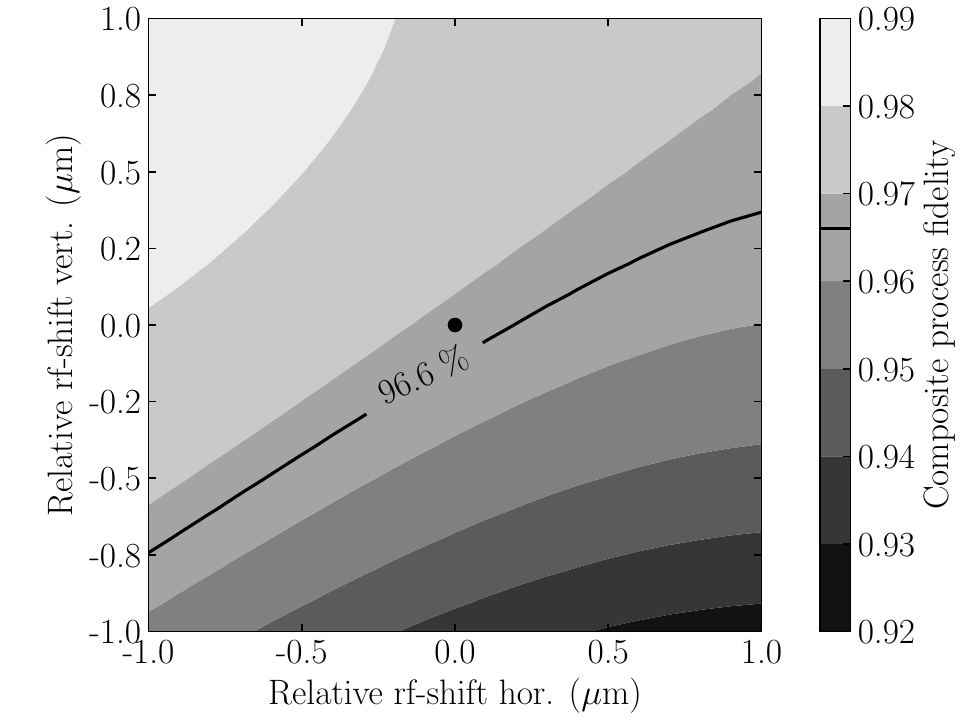}
	\caption{Result of simulating the realized cycle benchmarking circuits. The x and y axes are a small relative shift between the rf-null line and the measured micromotion gradient. The shifts are relative to the nominal results of our rf and microwave field simulations (black dot). The gray-scale shows the resulting composite process fidelity. The black solid line marks the 96.6\% value measured experimentally.}
	\label{fig:simulation}
\end{figure}

We now have an estimate for the involved spectroscopic shifts during individual-ion addressing, and simulate the outcome of applying these gates with the known imperfections. Combining this with the model of the two-qubit gate used in~\cite{zarantonello_robust_2019}, we can simulate the circuits done in the cycle benchmarking experiment, calculate the composite process fidelity and compare this to the experimental result. We get an estimated infidelity per dressed entangling gate of $96.8^{+0.8}_{-0.7}\%$ within the range uncertainties of relative rf-null position and microwave gradient field. The result of a wider scan over these parameters is shown in Fig. \ref{fig:simulation}. 

In future iterations, we can for example compensate this shift by applying a bichromatic single-ion addressing pulse (one sideband to cancel the shift and one sideband to address the ion), or by measuring the shifts individually and compensating for them by shifting the driving fields and applying a $\sigma_z$ gate on the non-addressed ion.

The uncompensated ac-Zeeman shift can also lead to a signature that is indistinguishable from crosstalk in the measurements shown in Fig. \ref{fig:crosstalk}. The observed ideal state is to have both ions return to the initially prepared $\ket{\uparrow\uparrow}$ state, as an even number of inversions is applied to the addressed ion and (ideally) identity is applied to the non-addressed ion. If a single applied pulse cannot reach the pole precisely, the randomization of rotation axes leads to a slow diffusion of the addressed ion's state off the pole of the Bloch sphere, leading to a signature that is, in this characterization, indistinguishable from cross-talk.  

\vspace{1em}

\bibliography{qc}

\end{document}